\documentclass{aa}  
\usepackage{graphicx}
\usepackage{txfonts}
\usepackage{graphicx}
\usepackage[normalem]{ulem}
\begin{document} 

\title{Bulge  RR~Lyrae  stars  in the  VVV  tile b201\thanks{Based  on observations  
    taken  within  the  ESO  VISTA  Public  Survey  VVV, Programme ID 179.B-2002.}}
   
\titlerunning{Bulge RR~Lyrae stars in the VVV tile b201}
\authorrunning{Gran et al.}

   %\subtitle{}

\author{
F.~Gran\inst{1}\fnmsep\inst{2}, 
D.~Minniti\inst{3}\fnmsep\inst{4}, 
R.~K.~Saito\inst{5}, 
C.~Navarrete\inst{1}\fnmsep\inst{2} 
I.~D\'ek\'any\inst{2}\fnmsep\inst{1}, 
I.~McDonald\inst{6}, 
R.~Contreras~Ramos\inst{1}\fnmsep\inst{2},
M.~Catelan\inst{1}\fnmsep\inst{2} 
}

\institute{ Instituto de Astrof\'isica, Pontificia Universidad Cat\'olica de Chile, 
Vicu\~na Mackenna 4860, Casilla 306, Santiago, Chile \\ \email{fgran@astro.puc.cl}
\and
Millennium Institute of Astrophysics (MAS), Santiago, Chile
\and
Departamento de Ciencias  Fisicas, Universidad Andr\'es Bello, Rep\'ublica 220, Santiago, Chile
\and
Vatican Observatory, V-00120 Vatican City State, Italy
\and  
Universidade   Federal   de   Sergipe,   Departamento   de   F\'isica, Av. Marechal Rondon s/n, 49100-000, S\~ao Crist\'ov\~ao, SE, Brazil
\and
Jordell   Bank  Centre   for  Astrophysics,   Alan   Turing  Building, Manchester, M13 9PL, UK
}

\date{Received June 04, 2014; accepted November 24, 2014}

% \abstract{}{}{}{}{} 
% 5 {} token are mandatory
 
  \abstract
{ The VISTA Variables in the V\'ia L\'actea (VVV) Survey is one of the six  ESO  public  surveys  currently  ongoing  at  the  VISTA
  telescope on  Cerro Paranal,  Chile.  VVV uses  near-IR ($ZYJHK_{\rm s}$) filters that at present provide photometry to a depth of  
  $K_{\rm  s} \sim 17.0$~mag  in up  to 36  epochs spanning  over four years, and aim at discovering more than  10$^6$ variable sources
  as well as trace the  structure  of the  Galactic  bulge and  part of  the southern disk. }
{  A  variability search  was  performed  to  find RR~Lyrae variable  stars. The  low  stellar density of  the VVV  tile
  \textit{b201},   which is centered  at   ($\ell,   b$)  $\sim$   ($-9^\circ, -9^\circ$),  makes it suitable  to  search  for
  variable stars. Previous studies have identified some RR~Lyrae stars using  optical  bands that  served  to  test  our search  procedure.
  The main  goal is to measure  the reddening, interstellar extinction, and distances of the RR~Lyrae stars and to study their distribution on the Milky Way bulge.  }
{  For each  star in  the  tile with  more than  25 epoch  ($\sim$90\% of the objects down to  $K_{\rm s} \sim 17.0$~mag), the standard deviation 
  and $\chi^2$  test were  calculated to  identify variable candidates. Periods were determined using the analysis  of variance.
  Objects with periods  in the  RR~Lyrae range of ${ 0.2  \leq P \leq  1.2 }$ days  were selected as candidate RR~Lyrae. They were individually 
  examined to exclude false positives. }
{ A total of  1.5 sq deg  were analyzed, and we found 39  RR~Lyr stars, 27 of which belong to the ab-type and 12 to the c-type. Our analysis 
  recovers all the previously  identified  RR~Lyrae variables in the field and  discovers 29 new RR~Lyr stars. The reddening and extinction toward 
  all the RRab stars in  this tile were derived, and distance estimations were obtained through  the  period--luminosity relation.   
  Despite the limited amount of RR~Lyrae stars studied, our results are consistent with a spheroidal or central distribution around $\sim 8.1$ and 
  $\sim 8.5$ kpc. for either the Cardelli or Nishiyama extinction law.  Our analysis does not reveal a stream-like structure.  
  Nevertheless, a larger area must be analyzed to definitively rule out streams. }
   {}


\maketitle

%________________________________________________________________%

\section{Introduction}
\label{sec:intro}

The  structure  and  evolution  of  the Galactic  bulge  is  far  from completely explored.  The high complexity of the Milky Way formation and the fact
that  we  are  embedded in  the  disk  of  the Galaxy,  impede  direct observations because of the extensive  layer of dust and  gas of the Galaxy in
this line of sight.  In the era of large-surveys, much effort have been  made to  reveal its structure,  taking advantage of  a wide
variety of near-, mid- and far-IR filters to bypass the layer of dust that  does  not  permit  deeper  observations  in  the  optical  bands
\citep[e.g.,][]{freudenreich96,skrutskie06,wright10}.

In  this context,  the VISTA  Variables  in the  V\'ia L\'actea  (VVV) Survey aims  to probe  the internal structure  of the Milky  Way bulge
through the near-IR  ($ZYJHK_{\rm{s}}$) filters, using as its main  distance  indicators  the  red-clump giants  and  pulsating
variable  stars  \citep{minniti10},   such as  RR~Lyrae  stars (hereafter RR~Lyr), classical Cepheids, anomalous  Cepheids and Miras,
and semiregular  variables. In  particular, RR~Lyr stars have been  useful probes of the bulge  structure and evolution
\citep{carney95,  gratton96,  mcwilliam10, soszynski11,  dekany13}. RR~Lyr   stars  are  old  and  low-mass   stars  \citep[$\approx  0.7
M_\odot$,][]{smith04} that are in  the horizontal branch (HB) stage (helium-burning  phase)  and  have  a  very  tight  period--luminosity
($P-L$) relation in  the near-IR bands \citep{longmore90, catelan04}. This relation is a powerful tool that has previously provided distances 
to   Galactic   globular    clusters   \citep{longmore90, delprincipe05,  delprincipe06, sollima06,  coppola11}, extragalactic
globular clusters, dSph galaxies  \citep{dallora04, pietrzynski08, borissova09}, and the  Galactic  center \citep{carney95, groenewegen08, dekany13}.
\cite{dekany13} searched  for near-IR counterparts of  the RR~Lyr stars that were found by the OGLE III survey \citep{soszynski11} in the Galactic bulge
in an area limited by $2^\circ \leq |b| \leq 7^\circ$ and $|\ell| \leq 10^\circ$. They derived the distance to the  Galactic center and discovered that the
RR~Lyr  trace a  more  spheroidal  shape  than the  red-clump  stars, which suggests that the Milky Way has a composite bulge.

In this paper, we have  selected a low density field to conduct a variability search.  We find new RR~Lyr stars
in  the VVV  tile \textit{b201},  which  allows us  to measure  precise reddening  and  distances to  the  bulge  in this  direction.
The   tile   \textit{b201} samples  the   spatial distribution   of    RR~Lyr   in   the    older   outer   bulge,
$b\gtrsim4^\circ$ away  from the  Galactic center region  analyzed by \cite{dekany13}. Our  motivation  also is to update  the
known  RR~Lyr  stars  present at  these  longitudes ($\ell \sim -10^\circ$) that are covered by the VVV Survey area.

\section{Observations}
\label{sec:obs}

The  VVV Survey is  being carried out with the 4m  Visible and Infrared Survey Telescope for
Astronomy (VISTA) located at ESO's Cerro Paranal Observatory in Chile, using the wide-field VISTA InfraRed Camera \citep[VIRCAM;][]{dalton06,
  emerson10}.    Photometric   catalogs   are based on   the   VISTA system, for which 2MASS \citep{cutri03}     coordinates     
  are   automatically  produced   by  the Cambridge Astronomical Survey Unit (CASU)\footnote{\url{http://casu.ast.cam.ac.uk/vistasp/}}.
    They   are publicly    available    through    the    VISTA    Science    Archive
(VSA)\footnote{\url{http://horus.roe.ac.uk/vsa/}}.   The  pixel  scale and field  of view (FoV) are 0.34\arcsec~pix$^{-1}$  and 1.64 deg$^2$.
This  FoV represents  one  ``tile''; the   entire  VVV observations  comprise  of 348  tiles  (196 covering  the bulge
and  152 and the  southern Galactic  plane). The VVV observation  schedule includes single-epoch near-IR photometry
in $ZYJHK_{\rm s}$ bands and a variability campaign in $K{}_{\rm s}$ with up to a total  of 100 epochs planned by the end
of    the    scheduled    observing   time    (1929    hours, \citealt{minniti10,  catelan11a}).  Technical  details  of the telescope and 
observation strategy can be found in \cite{minniti10} and \cite{saito12a}.

\begin{figure}
 \centering
 \includegraphics[scale=0.35]{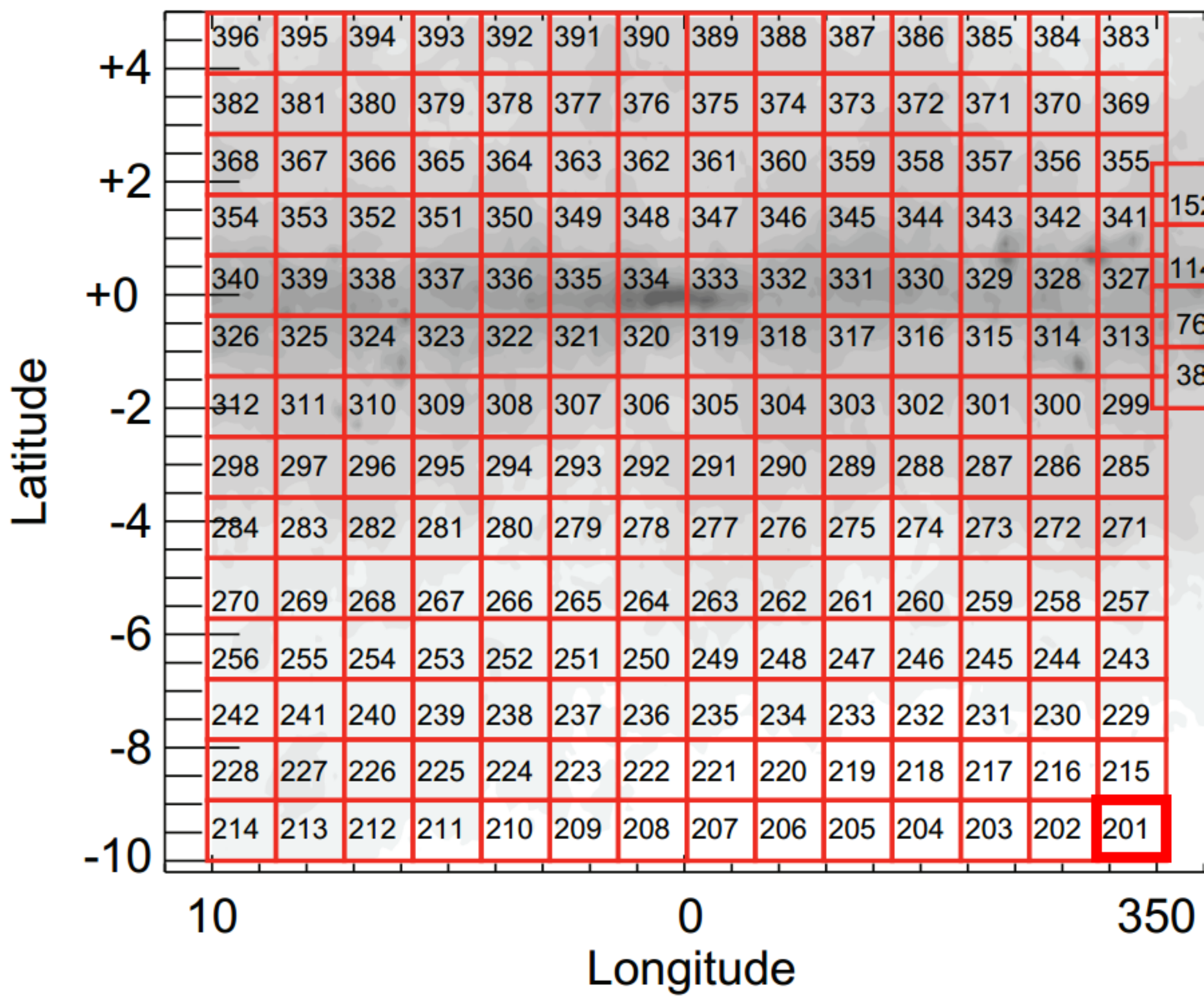}
 \caption{ VVV Survey bulge region, shown in Galactic coordinates with its individual numbered tiles.  Tile \textit{b201} 
 is highlighted in the bottom right corner. The Galactic center is contained in tile \textit{b333}. Adapted from \cite{catelan11a}. }
\label{fig:b201}
\end{figure}

In this study, we used observations of tile \textit{b201} ($\ell, b \approx -9^\circ,  -9^\circ$; Fig. \ref{fig:b201}  ).  
This tile is the bulge  field most distant from the Galactic center observed by the VVV, and  it has the lowest stellar density, as shown
in Fig. 2 of \cite{saito12b}. The VVV has observed this tile 36 times between April 2010  and July 2013.  A total  of 394\,047 point sources
with $K_{\rm s} \leq  17.0$~mag were detected, of which 353\,935  have more than 25 epochs of observations, which is the number we require to include
a source in this study.

The  main reason  to  choose this  tile  was its stellar density, lower than other observed VVV bulge tiles. Specifically, for a cutoff
of 25  points per light curve, there  were 353\,935 time series  to   be  analyzed in this tile  of   the  total  394\,047  ($\sim$   10  \%
excluded). This  number is similar to that of the  adjacent tile \textit{b202} with 408\,858 sources in  $K_{\rm{s}}$ band. Another reason for analying
this   tile   is   its   low  extinction   with   values   of $A_{K_{\rm{s}}}   \lesssim   0.1$~mag in certain lines of sight,   or   equivalently,
$A_{\rm{V}} \leq 0.35$~mag \citep{catelan11a,gonzalez11,gonzalez12}, which allowed the
previous      detection      of      RR~Lyr     stars      by \cite{swope42}, \cite{ponsen54} and \cite{kooreman66}. Even though it is not  completely free of  
selection biases,  the low  density implies  reduced crowding and a more accurate and deeper  photometry than for the other  fields in 
the VVV survey  area. Furthermore, the lower  extinction in this area minimizes the reddening uncertainties.

\subsection{Identifying and classifying variable sources}
\label{sec:id}

\begin{figure}
 \centering
 \includegraphics[scale=0.45]{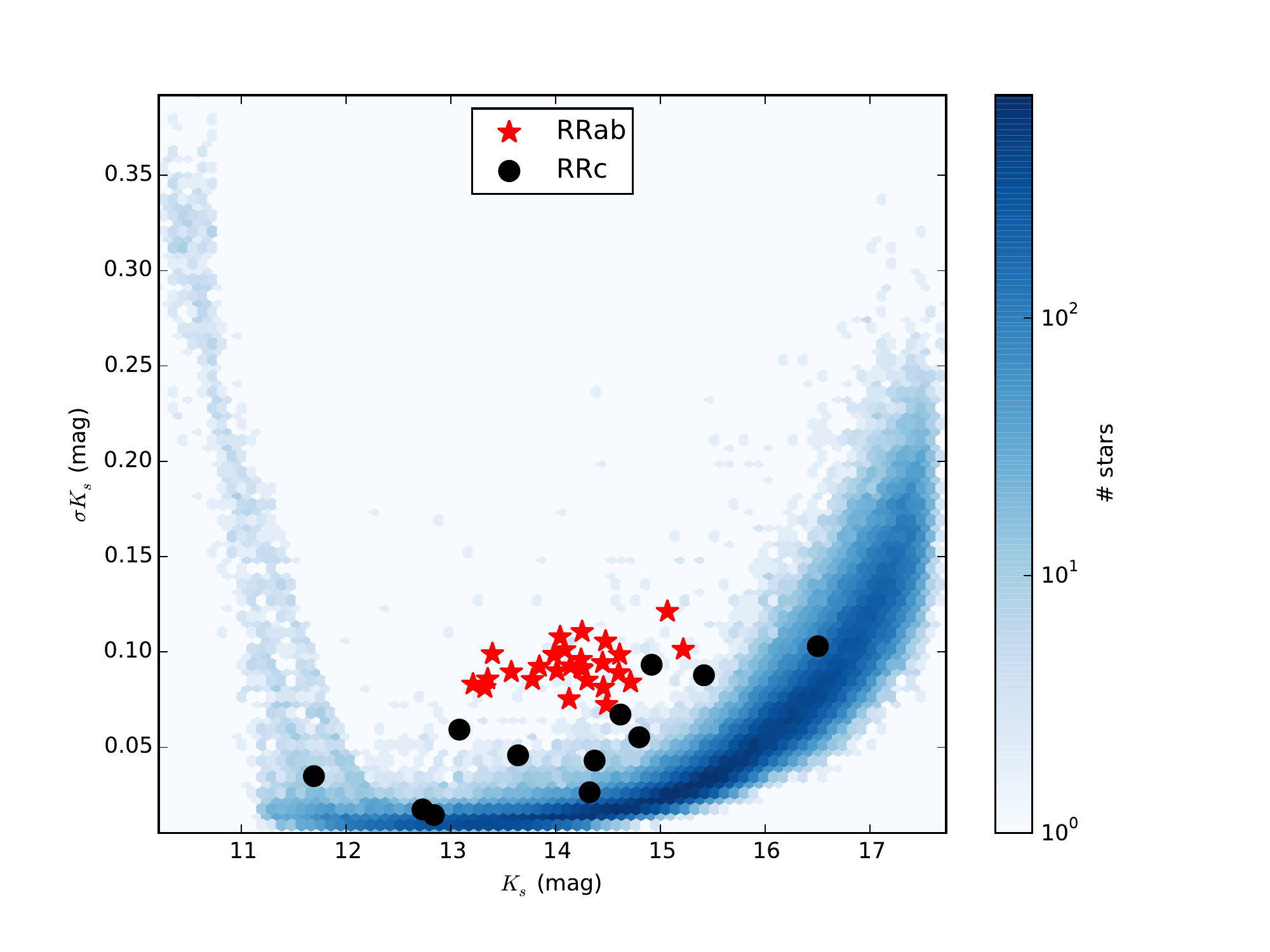}
 \caption{Magnitude-variability  diagram for  the sources  detected in \textit{b201}  with more  than  25 epochs  (levels  of blue).   The
   contribution of  saturated ($K_{\rm s}  \lesssim 12.0$~mag)  and faint ($K_{\rm  s} \gtrsim  16.5$~mag)  stars are  clearly  visible in  the
   $\sigma K_{\rm s}$ value. RRab are shown as red stars and RRc stars as black circles.}
\label{fig:Ksvar}
\end{figure}

A  semi-automated  classification scheme  was  developed  to select  the variable  stars.  First we  computed the  standard
deviation  ($\sigma  K_{\rm  s}$)  and  $\chi^2$ test  for  each  time series. The  cut-off for considering  a non-variable star  was determined
by simulating  a  random noise  $\chi^2$  distribution,  as described  by \cite{carpenter01}; this resulted in  $\chi^2 =  2$.  
This constraint places a lower limit of A$_{K_{\rm s}} \approx 0.03$~mag at $K_{\rm s} \lesssim 13$~mag on the amplitudes of the RR~Lyr to be detected.
We estimate, based in the $\omega$~Cen RRc stars K$_{\rm s}$-amplitude distribution (Navarrete  et al.  2014, subm.), 
that at K$_{\rm s} \sim 15.2$~mag our search procedure has a variability detection efficiency of $\sim$90\%, which drops to $\sim$50\% at K$_{\rm s} \sim 16.5$~mag. 
These candidates were analyzed by the analysis  of variance (AoV) statistis \citep{sczerny89} to determine their periods. 
Periods were restricted to the  range  in which  RR~Lyr  are typically found:  $0.2  \leq P {\rm (days)} \leq  1.2$. 
After  this step,  the light curves  of 167  candidates  were visually
inspected, and 42 objects were foundto be as RR~Lyr. Of the remaining 125  light curves, 32  were clearly  variables but  without a
clear variability class, and we classified the remaining  83 as  eclipsing-binary candidates.  Better  phase coverage  is needed to classify these variables.
Finally,  ten  variable-star  candidates  as selected  by  the algorithm  were determined  to  be non-variables  that  showed  period-aliasing  
of $\approx$1  day and  were therefore excluded from  further  analysis.   Fig.~\ref{fig:Ksvar}  presents the  magnitude--variability diagram  of 
all sources with more  than 25 epochs of observation, in  which the RR~Lyr
stars are identified.

\section{Results}
\subsection{Catalogue of RR~Lyr stars}

To characterize  our sample of  RR~Lyr  stars, Table \ref{tbl:summary} contains  the number of each  variable type found and the
mean, longest and shortest periods of each  population. These values define  the  Oosterhoff  and  Bailey  type of  the  RR~Lyr.
Additionally,  our RR~Lyr  candidates lie  in the  magnitude range  between  $13 \lesssim  $  $K_{\rm{s}}\,({\rm mag})  \lesssim
16.5$, resulting in a distance between $5.2 \lesssim d\,({\rm kpc}) \lesssim 13.6$.

\begin{figure}
 \centering
 \includegraphics[scale=0.55]{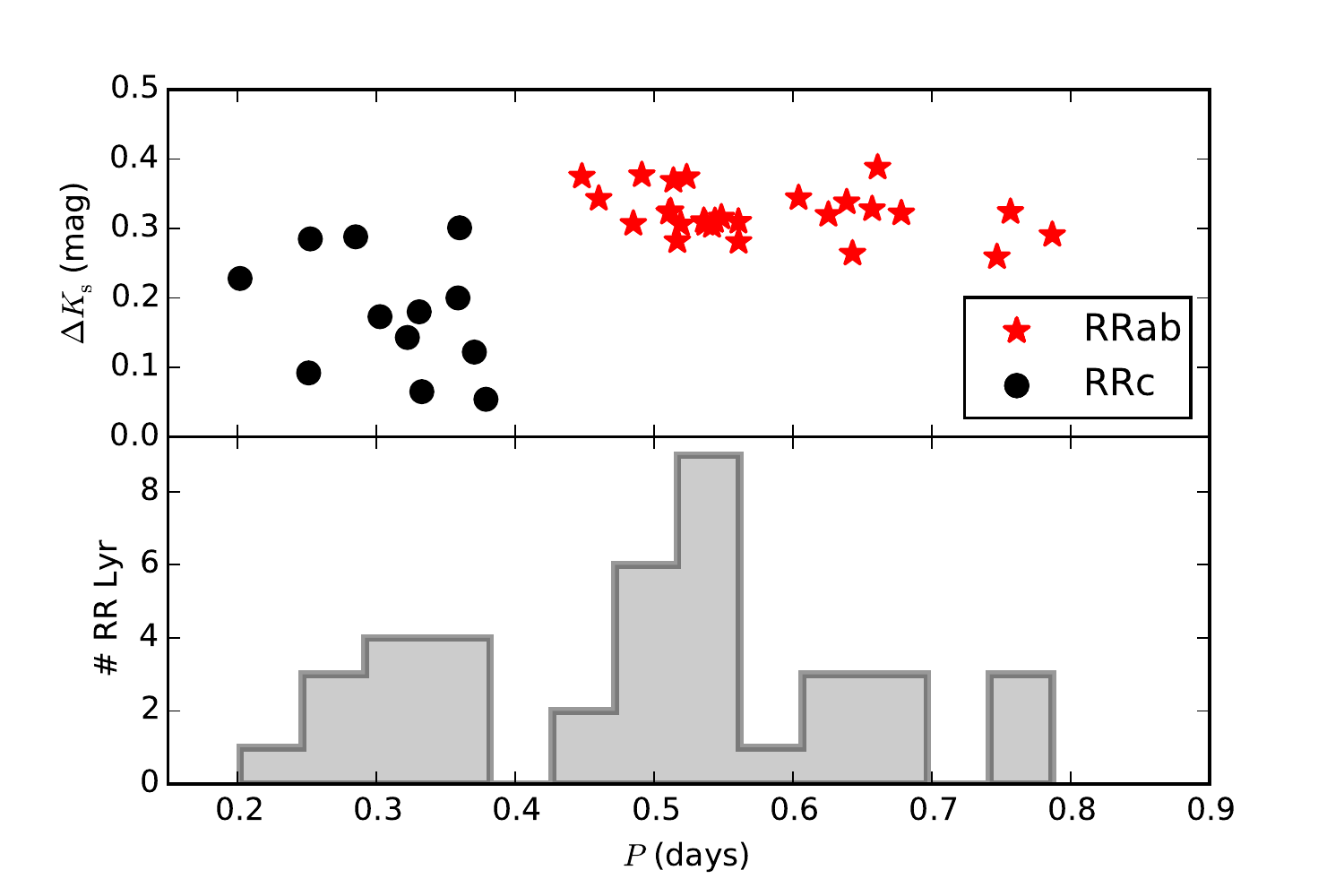}
 \caption{{\bf Top}: Bailey  diagram for  the 39 RR~Lyr variable  stars. As expected,  there is a  clear separation  between the  amplitude and
   period  among  RRab  (upper  right)  and RRc  stars  (lower  left), following   the    same   symbols   as in  Fig.~\ref{fig:Ksvar}.
   {\bf Bottom}:  Period  histogram of  the  RR~Lyr.}
\label{fig:Bailey} 
\end{figure}

\cite{bailey02}  classified  the RR~Lyr  (in  that  epoch  called ``cluster variables'') in  three groups, the a-, b-  and c-type by its
period and  light-curve shape (amplitude of variation).  The first two groups were merged  later into the ab-type known today. This relation
of period  and amplitude (today called  the Bailey diagram) is  shown in the upper panel of Fig.~\ref{fig:Bailey}.

\begin{table}
\centering
\caption{Summary  of the  periods (in  days)  of the  RR~Lyr found  in \textit{b201}.}
\begin{tabular}{c|c|c|c|c}
\hline\hline
  Variability type   & Quantity   & $\langle P \rangle$ & $P_{\rm{min}}$ & $P_{\rm{max}}$ \\
                     &            &        (days)       &     (days)     &     (days)     \\
\hline\hline
  RRab               & 27         & 0.578               & 0.448          & 0.787          \\
  RRc                & 12         & 0.312               & 0.202          & 0.379          \\
\hline\hline
\end{tabular}
\label{tbl:summary}
\end{table}

\begin{figure}
\centering
\includegraphics[scale=0.5]{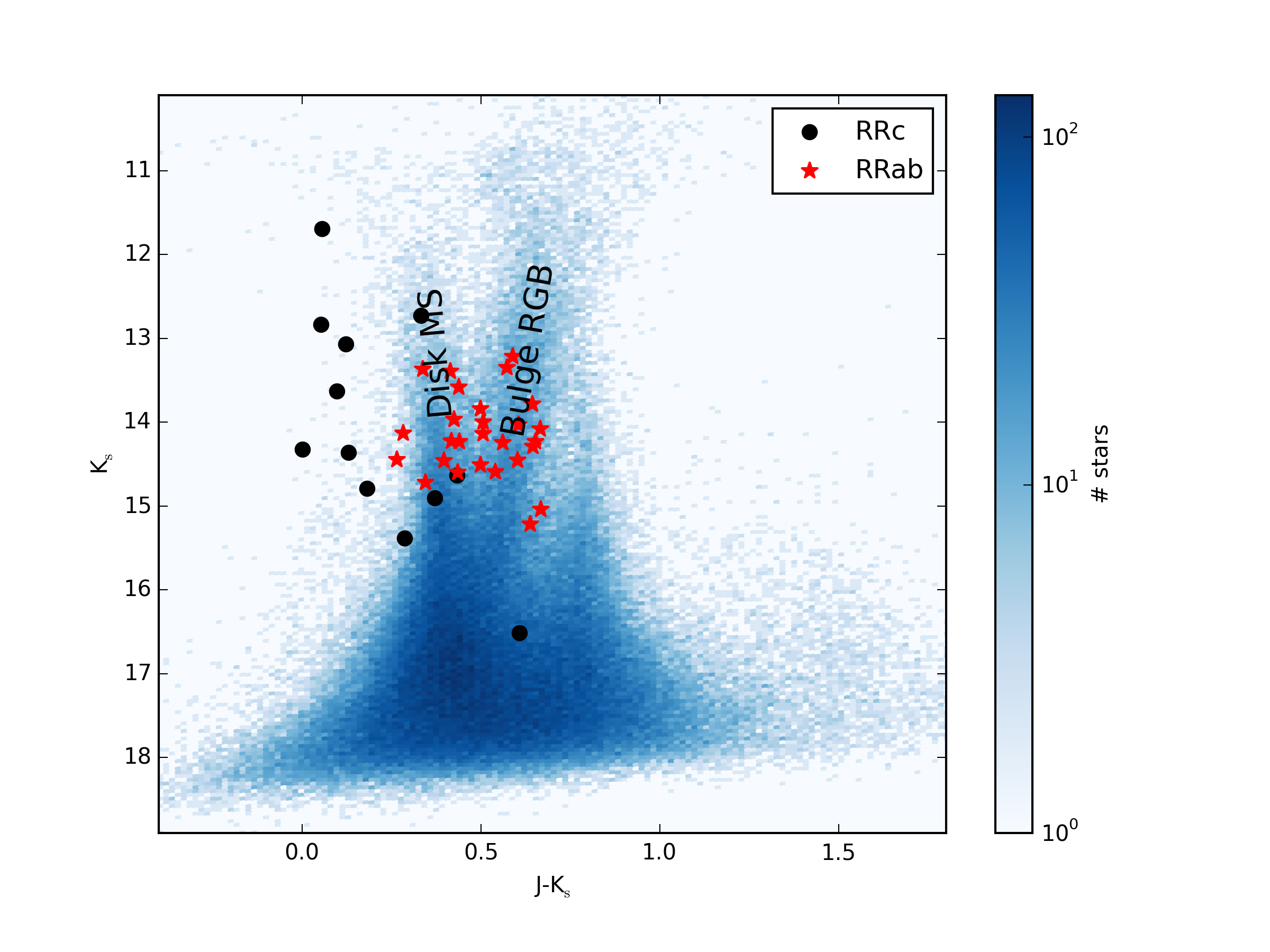}
\caption{Hess diagram of the tile \textit{b201}. The disk MS and bulge RGB are easily recognizable from  this field and similar to the ones
  derived in \cite{saito12b}.  The symbols of the RR~Lyr are the same as in Fig.~\ref{fig:Ksvar}.}
\label{fig:cmd}
\end{figure}

$J$ and $H$  photometry taken  during 2010 and the $K_{\rm{s}}$-mean magnitudes are  shown in  Table \ref{tbl:colors}.  A brief discussion  of the
error   introduced  in   the  RR~Lyr   color   ($J-K_{\rm{s}}$)  and ($H-K_{\rm{s}}$)   is  presented   in   Sect.~\ref{sec:reddening}.
The RR~Lyr stars are  shown on a Hess color--magnitude diagram of  the whole tile  in Fig.~\ref{fig:cmd}.  The  bulge red-giant
branch (RGB) and the disk main-sequence (MS) stars are presented in the Hess diagram, with the  label aligned in the direction  of each branch. The
superposition of a  young population (the disk) hides  the MS turn-off and  part  of  the  Galactic  bulge  HB,  where  most  of  our  RR~Lyr
belong. However, the bulge instability strip locus can be inferred over the disk  MS in  a color range  between ${0.1  \lesssim (J-K_{\rm{s}})
\lesssim  0.6}$, according  to the  position  of the  RR~Lyr in  the diagram.

\begin{table*}
\centering
\caption{Color information  of RRab  and RRc stars  found in  tile \textit{b201}.  We list the   VVV ID,  right  ascension  (J2000),
  declination  (J2000),  magnitude-weighted mean  from the Fourier  fit  $\langle K_{\rm{s}}  \rangle$,   and  ($J-K_{\rm{s}}$)  and  ($H-K_{\rm{s}}$)
  colors from the first K$_{\rm{s}}$-band epoch.}
\begin{tabular}{c|c|c|c|c|c}
\hline\hline
VVV ID (RRab) & RA(J2000) & DEC(J2000) & $\langle$ $K_{\rm{s}} \rangle$ & $(J-K_{\rm{s}})$\tablefootmark{a} & $(H-K_{\rm{s}})$\tablefootmark{a} \\
\hline\hline                       
VVV J2712426.16-421241.6 & 18:05:37.7 & $-$42:12:41.6 & 14.59 & 0.54 & 0.22 \\
VVV J2712711.57-421314.0 & 18:05:48.8 & $-$42:13:14.0 & 14.23 & 0.65 & 0.29 \\
VVV J2712912.05-420223.7 & 18:05:56.8 & $-$42:02:23.7 & 14.23 & 0.44 & 0.17 \\
VVV J2711512.06-415330.1 & 18:05:00.8 & $-$41:53:30.1 & 14.00 & 0.51 & 0.22 \\
VVV J2703434.23-413322.1 & 18:02:18.3 & $-$41:33:22.1 & 13.35 & 0.57 & 0.24 \\
VVV J2703536.01-412829.4 & 18:02:22.4 & $-$41:28:29.4 & 13.97 & 0.42 & 0.20 \\
VVV J2703343.00-412731.0 & 18:02:14.9 & $-$41:27:31.0 & 14.45 & 0.60 & 0.25 \\
VVV J2711523.24-413619.8 & 18:05:01.5 & $-$41:36:19.8 & 13.78 & 0.64 & 0.27 \\
VVV J2710804.03-413241.7 & 18:04:32.3 & $-$41:32:41.7 & 14.51 & 0.50 & 0.21 \\
VVV J2704944.51-412248.6 & 18:03:19.0 & $-$41:22:48.6 & 13.36 & 0.34 & 0.16 \\
VVV J2714215.15-414058.6 & 18:06:49.0 & $-$41:40:58.6 & 14.29 & 0.64 & 0.25 \\
VVV J2710759.32-411700.5 & 18:04:31.9 & $-$41:17:00.5 & 13.39 & 0.41 & 0.20 \\
VVV J2714823.59-413121.7 & 18:07:13.6 & $-$41:31:21.7 & 14.44 & 0.26 & 0.15 \\
VVV J2711644.06-411744.6 & 18:05:06.9 & $-$41:17:44.6 & 13.84 & 0.50 & 0.21 \\
VVV J2710158.99-410727.3 & 18:04:07.9 & $-$41:07:27.3 & 14.22 & 0.42 & 0.20 \\
VVV J2704656.26-421805.3  & 18:03:07.7 & $-$42:18:05.3 & 15.22 & 0.64 & 0.26 \\
VVV J2701149.18-420358.3  & 18:00:47.3 & $-$42:03:58.3 & 13.22 & 0.59 & 0.25 \\
VVV J2705101.04-405300.9 & 18:03:24.1 & $-$40:53:00.9 & 13.58 & 0.44 & 0.18 \\
VVV J2711836.69-405946.8 & 18:05:14.4 & $-$40:59:46.8 & 14.24 & 0.56 & 0.25 \\
VVV J2705334.00-421303.7  & 18:03:34.3 & $-$42:13:03.7 & 15.04 & 0.67 & 0.28 \\
VVV J2702317.06-420003.9  & 18:01:33.1 & $-$42:00:03.9 & 14.14 & 0.50 & 0.23 \\
VVV J2705622.48-420853.7  & 18:03:45.5 & $-$42:08:53.7 & 14.08 & 0.66 & 0.29 \\
VVV J2711018.90-421253.4  & 18:04:41.2 & $-$42:12:53.4 & 14.03 & 0.60 & 0.27 \\
VVV J2712342.04-421647.8  & 18:05:34.8 & $-$42:16:47.8 & 14.13 & 0.28 & 0.15 \\
VVV J2712855.63-421832.3  & 18:05:55.7 & $-$42:18:32.3 & 14.46 & 0.40 &  $-$ \tablefootmark{b} \\
VVV J2710638.97-420755.7  & 18:04:26.6 & $-$42:07:55.7 & 14.59 & 0.43 & 0.19 \\
VVV J2702400.61-415026.5  & 18:01:36.0 & $-$41:50:26.5 & 14.72 & 0.34 & 0.18 \\ [0.5ex]
\hline\hline
VVV ID (RRc) & RA(J2000) & DEC(J2000) & $\langle$ $K_{\rm{s}} \rangle$ & $(J-K_{\rm{s}})$ & $(H-K_{\rm{s}})$ \\ 
\hline\hline 
VVV J2711026.32-415037.7 & 18:04:41.7 & $-$41:50:37.7 & 15.39 & 0.29 & 0.13 \\
VVV J2702945.91-413239.9 & 18:01:59.0 & $-$41:32:39.9 & 12.73 & 0.33 & 0.09 \\
VVV J2710200.43-413446.0 & 18:04:08.0 & $-$41:34:46.0 & 12.84 & 0.05 & 0.02 \\
VVV J2705846.71-411723.2 & 18:03:55.1 & $-$41:17:23.2 & 14.63 & 0.43 & 0.17 \\
VVV J2705404.77-422910.0 & 18:03:36.3 & $-$42:29:10.0 & 14.36 & 0.13 & 0.08 \\
VVV J2713458.02-412608.5 & 18:06:19.9 & $-$41:26:08.5 & 13.63 & 0.10 & 0.06 \\
VVV J2711142.24-411611.0 & 18:04:46.8 & $-$41:16:11.0 & 14.79 & 0.18 & 0.10 \\
VVV J2714054.14-412637.6 & 18:06:43.6 & $-$41:26:37.6 & 14.33 & 0.01 & 0.02 \\
VVV J2711803.26-411635.8 & 18:05:12.2 & $-$41:16:35.8 & 11.70 & 0.05 & 0.01 \\
VVV J2710857.12-410712.8 & 18:04:35.8 & $-$41:07:12.8 & 14.90 & 0.37 & 0.17 \\
VVV J2702329.86-415953.8  & 18:01:34.0 & $-$41:59:53.8 & 16.51 & 0.61 & 0.22 \\
VVV J2704051.15-420503.6  & 18:02:43.4 & $-$42:05:03.6 & 13.07 & 0.12 & 0.10 \\ [0.5ex]
\hline\hline
\end{tabular}
\tablefoot{
\tablefoottext{a}{ Typical ($J-H$) and ($H-K_{\rm{s}}$) photometrical errors  are $\approx$ 0.05~mag.}
\tablefoottext{b}{$H$-magnitude not available. } }
\label{tbl:colors}
\end{table*}

The period distribution of the RRab and RRc stars is divided by a gap  in $P  \sim 0.4$  days, as  shown in  the bottom  panel  of Fig.~\ref{fig:Bailey}, 
which reflects the  distribution  derived  by \cite{kunder09}  in  $V$  and  $R$ bands.  These authors analyzed  the
Oosterhoff  type of  RR~Lyr  in the  Galactic bulge.   This near-IR  diagram  shows  a  different  slope   for  the  RRab  in  the  optical
wavelengths. It is more horizontal,   similar   to  that   reported   by \cite{gavrilchenko14} and Navarrete et  al.  (2014, subm.)  using mid-
and near-IR data.

\subsection{Previously known RR~Lyr stars}

As a test of consistency, we compared our RR~Lyr positions and periods with those found in the literature. There are already ten variables in
the International Variable Star Index\footnote{\url{http://www.aavso.org/vsx}} (VSX) catalog classified as RR~Lyr in the analyzed area.

We recovered  all  ten previously  known  variables and  correctly   classified them  as RR~Lyr.  These  ten RR~Lyr  are listed  in Table
\ref{tbl:VSX}, along  with their  General Catalogue of  Variable Stars \citep[GCVS;][]{samus09}   designations,   periods,  and   astrometric
offsets.  In  three cases  (MO~CrA, V397~CrA and  V467~CrA),  periods different from those in  the  literature  were  identified. MO  CrA  was
classified, without  an assigned period, as a  ``cluster variable'' by \cite{swope42}.   V397~CrA  and V467~CrA  were  previously observed  by
\cite{ponsen54}    and    \cite{kooreman66},    respectively.  \cite{ponsen54}  classified V397~CrA as  an RRc,  but with  only 16
points in the  light curve. \cite{kooreman66} classified V467~CrA as an  RRab, remarking  that this  star was very  faint, with  a large
dispersion  in magnitude  in the  120 available  plates. V475  CrA and V493~CrA do not have counterparts within our initial 2\arcsec matching
radius,  but RR~Lyr  with near  identical periods  were  identified at 17\arcsec and 31\arcsec, respectively,  from their VSX positions. 
According to CASU, VVV astrometry is accurate to $\approx$50 mas and   the  similarity in  periods make  it very  likely that  V475~CrA and
V493~CrA are counterparts to VVV J2705622.48-420853.7 and VVV J2714823.59-413121.7.

\begin{table*}
\centering
\caption{Results of  the match with the  VVV tile \textit{b201} data and the VSX catalog. }
\begin{tabular}{c|c|c|c|c}
\hline\hline
{\bf VVV ID} & GCVS & $P_{\rm{VVV}}$ & $P_{\rm{Lit.}}$ &  $d$   \\
   &      &      (days)    &      (days)     & ($\arcsec$) \\
\hline\hline
  VVV J2705101.04-405300.9 & MO CrA   & 0.657005 & --        & 0.475 \\
  VVV J2712855.63-421832.3 & V397 CrA & 0.491175 & 0.3293815 & 0.771 \\
  VVV J2701149.18-420358.3 & V463 CrA & 0.604052 & 0.6040585 & 0.139 \\
  VVV J2702329.86-415953.8 & V467 CrA & 0.359989 & 0.4480160 & 1.931 \\
  VVV J2703434.23-413322.1 & V469 CrA & 0.678028 & 0.6780236 & 0.528 \\
  VVV J2705622.48-420853.7 & V475 CrA & 0.511933 & 0.5119430 & 17.41 \\
  VVV J2710759.32-411700.5 & V482 CrA & 0.541709 & 0.5417140 & 2.725 \\
  VVV J2712342.04-421647.8 & V483 CrA & 0.485059 & 0.4850490 & 0.105 \\
  VVV J2712711.57-421314.0 & V486 CrA & 0.460161 & 0.4601694 & 0.573 \\
  VVV J2714823.59-413121.7 & V493 CrA & 0.519262 & 0.5194291 & 31.26 \\
\hline\hline
\end{tabular}
\label{tbl:VSX}
\end{table*} 

\subsection{RR~Lyr star with the shortest period in our catalog}
\label{shortest}

We found that  the RRc  candidate VVV J2705846.71-411723.2 has a very short period of $P = 0.202$~days.  In contrast, the RRc with the shortest period  found
in the OGLE III sample has $P=0.237$~days \citep{soszynski11}.  This variable warrants a more detailed study to confirm that it is an RRc instead
of an eclipsing binary or a long-period SX Phe, as reported by \cite{cohen12}.

\subsection{Fourier coefficients}
\label{metallicity}

\begin{figure*}
 \centering
 \includegraphics[scale=0.42]{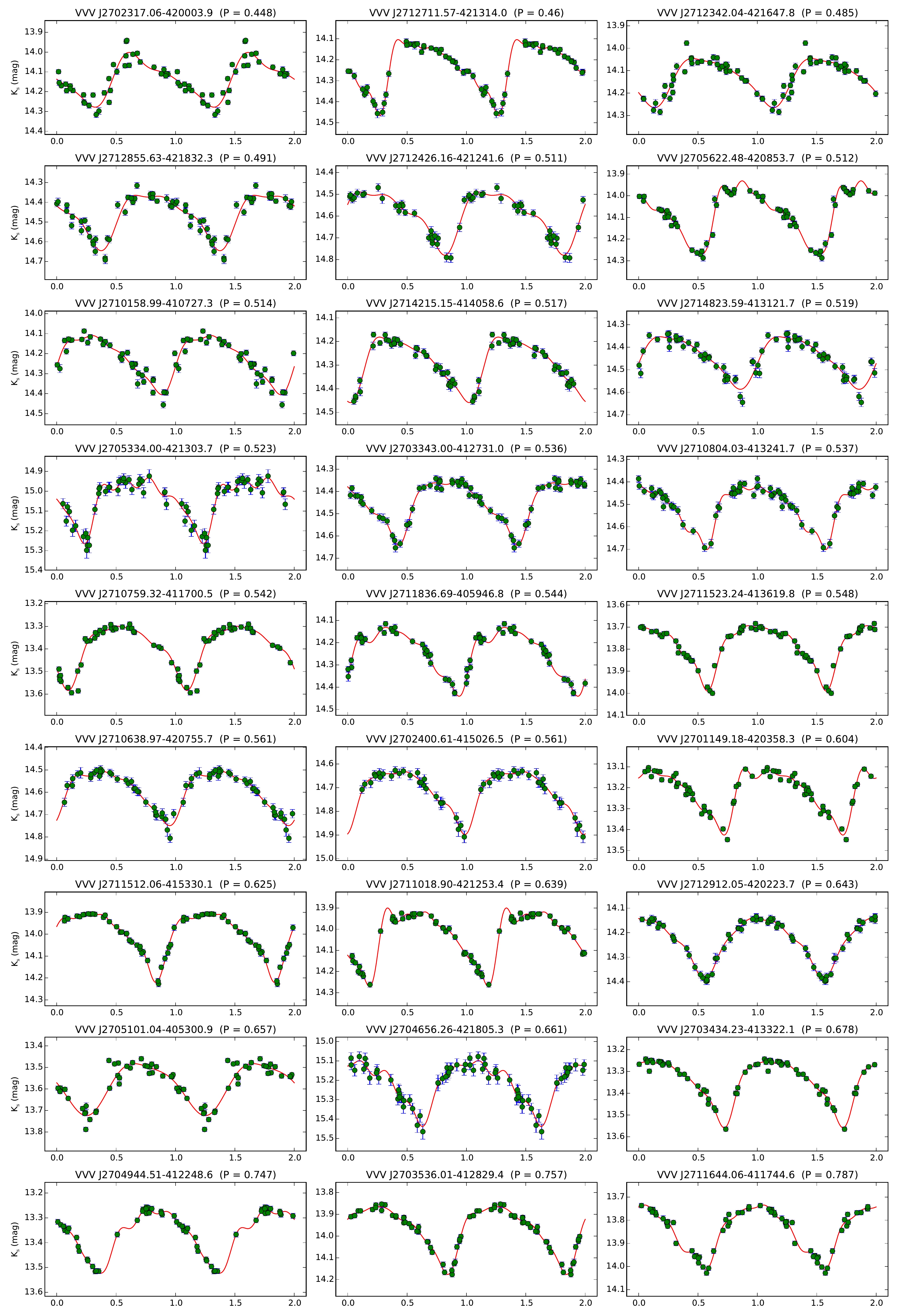}
 \label{RRab}
 \caption{30  RRab stars found in  tile \textit{b201}, sorted  by increasing period. The solid  line represents  the Fourier
   decomposition of  each RR~Lyr using  the DFF method.   The internal name  and period  are indicated  at  the top  of each  light-curve.
   Error bars are  plotted, but  are generally smaller than the point sizes. }
\label{fig:RRab}
\end{figure*}

\begin{figure*}
 \centering
 \includegraphics[scale=0.42]{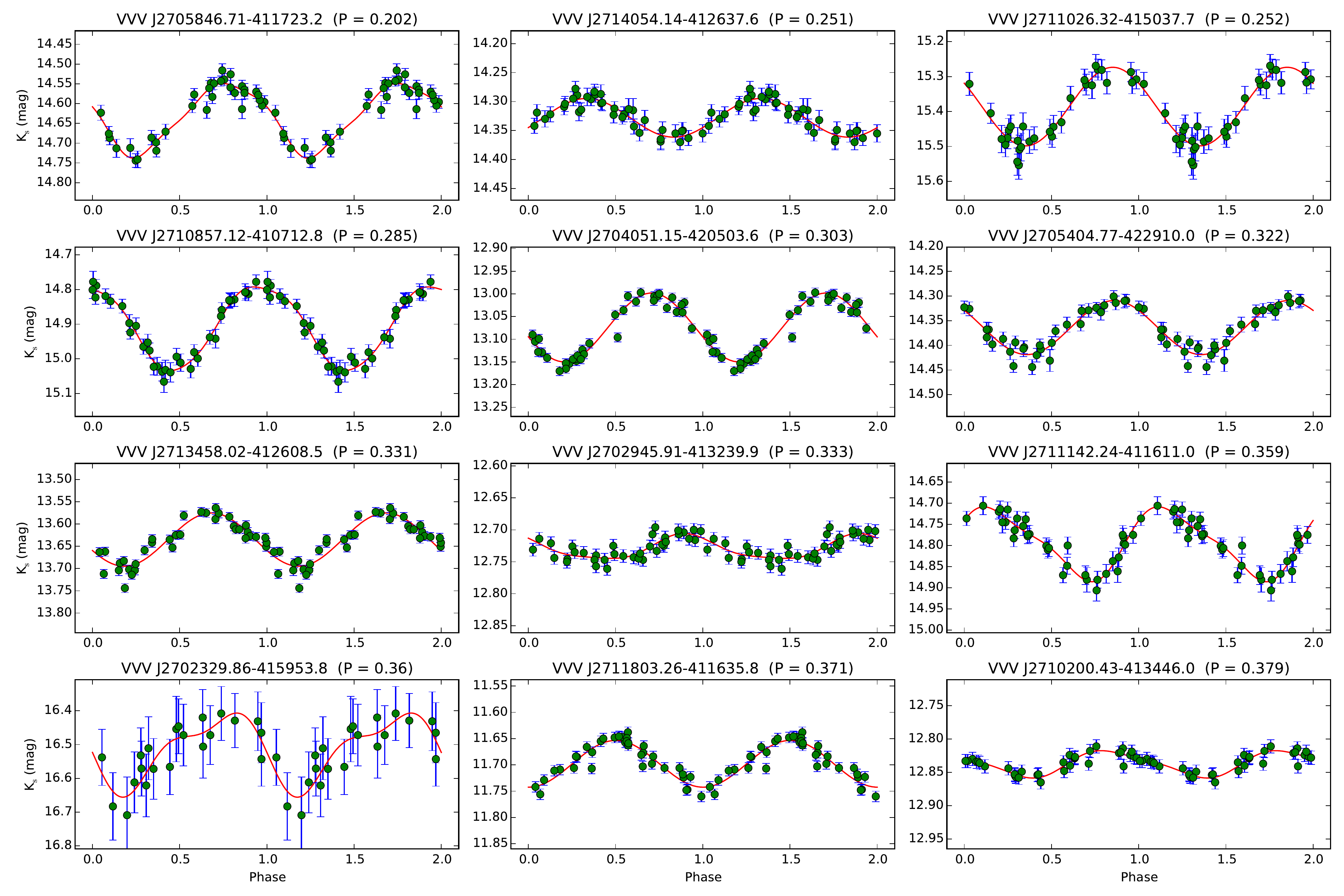}
 \caption{Same as Fig.~\ref{fig:RRab}, but for the RRc stars.}
\label{fig:RRc}
\end{figure*}

The   complete  catalog   of  light curves   is  shown   in  Figs. \ref{fig:RRab}  and   \ref{fig:RRc}  for  the  RRab   and  RRc  types.
Following   an approach similar    to  that  of \cite{dekany13}, the light curves of the RRab and RRc stars were fitted
with  Fourier   series  compound   of  sines,  up   to  sixth order using  the direct  Fourier fitting  (DFF)
method of \cite{kovacs07}. This order was the highest possible calculated, but in some cases the returned order was lower to prevent over-fitting of the light-curve.
This method  allows us to recover the mean $K_{\rm{s}}$-magnitude   plus  the  Fourier   coefficients  ($A_{21}$,
$\phi_{21}$, ..., $A_{61}$, $\phi_{61}$), which are shown in Table \ref{tbl:Fourier}.

The photometric  iron abundance  of RR Lyr  stars is  typically found by    analyzyng     of    $V$-    and     $I$-band    light-curves
\citep[e.g.,][]{jurcsik96,  smolec05}.   However,  since there  is  no relationship between  the Fourier parameters and  iron content already
established in  the near-IR, it cannot be  determined individually for our variables. Hence, we  used the value of \cite{pietrukowicz12}, who
found that RR~Lyr trace the distribution of  the metallicity in the Galactic bulge with  a sharply peaked distribution, centered
on  $-1.02$~dex  with   a  dispersion  of  0.25~dex.  This metallicity complete agree with the spectroscopic value 
found by \cite{walker91} of $\langle [Fe/H] \rangle = -1.0$~dex with a dispersion of $0.16$~dex. They analyzed 59 RR~Lyr in Baade's window.
Taking a fixed metallicity  does not  significantly affect  the  distance estimation; it imparts a  mean of $0.20$~kpc ($\sim 3$\% at 8~kpc)  uncertainty 
to distances derived  using the $P-L$  relation, taking into account the extreme values on the dispersion of [Fe/H].

\begin{table*}
\centering
\caption{Period and Fourier coefficients for the RR Lyr in tile \textit{b201}.}  \scalebox{0.8}{
\begin{tabular}{c|c|c|c|c|c|c|c|c|c|c|c|c|c}
\hline\hline
VVV ID (RRab) & $P$ (days) & $A_1$ & $\phi_{1}$ & $A_{21}$ & $\phi_{21}$ & $A_{31}$ & $\phi_{31}$ & $A_{41}$ & $\phi_{41}$ & $A_{51}$ & $\phi_{51}$ & $A_{61}$ & $\phi_{61}$ \\
\hline\hline

VVV J2702317.06-420003.9 & 0.448043 & 0.108 & 0.146 & 0.442 &  3.361 & 0.156 &  0.038  &  $-$  &  $-$    &  $-$  &  $-$    &  $-$  &  $-$    \\
VVV J2712711.57-421314.0 & 0.460161 & 0.127 & 0.595 & 0.454 &  3.768 & 0.336 &  1.103  & 0.179 & $-$1.599  & 0.156 &  1.955  & 0.132 & $-$0.353  \\
VVV J2712342.04-421647.8 & 0.485059 & 0.098 & 1.042 & 0.336 &  3.942 & 0.109 &  1.388  &  $-$  &  $-$    &  $-$  &  $-$    &  $-$  &  $-$    \\
VVV J2712855.63-421832.3 & 0.491175 & 0.122 & 5.804 & 0.376 & $-$8.542 & 0.196 & $-$17.03  &  $-$  &  $-$    &  $-$  &  $-$    &  $-$  &  $-$    \\
VVV J2712426.16-421241.6 & 0.510823 & 0.120 & 3.069 & 0.403 & $-$2.386 & 0.228 & $-$4.592  &  $-$  &  $-$    &  $-$  &  $-$    &  $-$  &  $-$    \\
VVV J2705622.48-420853.7 & 0.511933 & 0.137 & 5.191 & 0.413 & $-$8.957 & 0.174 & $-$11.045 & 0.123 & $-$20.52  & 0.117 & $-$23.676 & 0.099 & $-$27.225 \\
VVV J2710158.99-410727.3 & 0.513884 & 0.128 & 2.769 & 0.295 & $-$2.487 & 0.157 & $-$4.873  & 0.116 & $-$7.579  &  $-$  & $-$13.846 &  $-$  &  $-$    \\
VVV J2714215.15-414058.6 & 0.516757 & 0.116 & 1.954 & 0.393 & $-$2.456 & 0.183 & $-$5.302  & 0.098 & $-$1.648  &  $-$  & $-$9.772  &  $-$  &  $-$    \\
VVV J2714823.59-413121.7 & 0.519262 & 0.105 & 2.818 & 0.317 & $-$2.442 & 0.118 & $-$4.912  &  $-$  &  $-$    &  $-$  &  $-$    &  $-$  &  $-$    \\
VVV J2705334.00-421303.7 & 0.523466 & 0.118 & 0.447 & 0.432 & $ $4.396 & 0.310 &  2.112  & 0.175 &  0.423  & 0.197 &  3.431  &  $-$  &  $-$    \\
VVV J2703343.00-412731.0 & 0.535809 & 0.117 & 5.564 & 0.321 & $-$8.513 & 0.258 & $-$11.026 & 0.102 & $-$19.275 & 0.072 & $-$27.465 &  $-$  &  $-$    \\
VVV J2710804.03-413241.7 & 0.536988 & 0.114 & 4.762 & 0.355 & $-$8.314 & 0.159 & $-$10.869 & 0.141 & $-$13.575 & 0.131 & $-$21.91  & 0.121 & $-$24.132 \\
VVV J2710759.32-411700.5 & 0.541709 & 0.113 & 1.205 & 0.378 & $-$2.069 & 0.254 &  2.238  & 0.111 &  0.374  &  $-$  &  $-$    &  $-$  &  $-$    \\
VVV J2711836.69-405946.8 & 0.543833 & 0.126 & 2.396 & 0.290 & $-$1.936 & 0.173 & $-$4.095  & 0.187 & $-$7.041  & 0.125 & $-$9.401  &  $-$  &  $-$    \\
VVV J2711523.24-413619.8 & 0.548459 & 0.113 & 4.514 & 0.407 & $-$8.315 & 0.185 & $-$10.374 & 0.177 & $-$12.632 & 0.123 & $-$20.615 &  $-$  &  $-$    \\
VVV J2710638.97-420755.7 & 0.560748 & 0.107 & 2.260 & 0.330 & $-$2.067 & 0.159 & $-$4.570  & 0.114 & $-$7.347  &  $-$  &  $-$    &  $-$  &  $-$    \\
VVV J2702400.61-415026.5 & 0.560889 & 0.108 & 2.005 & 0.325 & $-$2.066 & 0.239 & $-$4.529  & 0.131 & $-$6.443  & 0.069 & $-$8.248  &  $-$  &  $-$    \\
VVV J2701149.18-420358.3 & 0.604052 & 0.122 & 3.905 & 0.387 & $-$2.170 & 0.255 & $-$11.105 & 0.235 & $-$13.759 & 0.131 & $-$16.498 &  $-$  &  $-$    \\
VVV J2711512.06-415330.1 & 0.625496 & 0.122 & 2.897 & 0.375 & $-$2.041 & 0.271 & $-$4.298  & 0.130 & $-$6.362  & 0.055 & $-$14.464 & 0.055 & $-$15.268 \\
VVV J2711018.90-421253.4 & 0.638694 & 0.136 & 1.014 & 0.404 &  4.155 & 0.304 &  1.629  & 0.218 & $-$1.227  & 0.172 & $-$3.558  & 0.089 & $-$0.497  \\
VVV J2712912.05-420223.7 & 0.642877 & 0.109 & 4.470 & 0.266 & $-$8.214 & 0.130 & $-$9.892  & 0.091 & $-$17.681 &  $-$  &  $-$    &  $-$  &  $-$    \\
VVV J2705101.04-405300.9 & 0.657015 & 0.116 & 0.131 & 0.238 &  4.176 &  $-$  &  $-$    &  $-$  &  $-$    &  $-$  &  $-$    &  $-$  &  $-$    \\
VVV J2704656.26-421805.3 & 0.660970 & 0.133 & 4.211 & 0.366 & $-$8.292 & 0.152 & $-$10.250 & 0.137 & $-$13.372 & 0.144 & $-$20.435 &  $-$  &  $-$    \\
VVV J2703434.23-413322.1 & 0.678028 & 0.121 & 3.676 & 0.386 & $-$2.160 & 0.234 & $-$10.655 & 0.114 & $-$12.769 & 0.074 & $-$14.812 &  $-$  &  $-$    \\
VVV J2704944.51-412248.6 & 0.746910 & 0.113 & 5.754 & 0.241 & $-$7.914 & 0.126 & $-$16.045 & 0.102 & $-$18.418 & 0.104 & $-$28.086 &  $-$  &  $-$    \\
VVV J2703536.01-412829.4 & 0.756627 & 0.126 & 2.947 & 0.372 & $-$2.210 & 0.214 & $-$4.254  & 0.124 & $-$6.989  & 0.063 & $-$8.932  &  $-$  &  $-$    \\
VVV J2711644.06-411744.6 & 0.786726 & 0.122 & 4.691 & 0.207 & $-$8.180 & 0.119 & $-$10.891 & 0.114 & $-$13.366 & 0.110 & $-$21.196 &  $-$  &  $-$    \\ [0.5ex]
\hline\hline
\end{tabular}
}
\scalebox{0.9}{
\begin{tabular}{c|c|c|c|c|c|c|c}
\hline\hline
VVV ID (RRc) & $P$ (days) & $A_1$ & $\phi_{1}$ & $A_{21}$ & $\phi_{21}$ & $A_{31}$& $\phi_{31}$ \\
\hline\hline
VVV J2705846.71-411723.2 & 0.201895 & 0.089 & 6.168 & 0.109 & $-$7.217 & 0.101 & $-$14.489  \\
VVV J2714054.14-412637.6 & 0.251285 & 0.033 & 2.613 &  $-$  &  $-$   &  $-$  &  $-$     \\
VVV J2711026.32-415037.7 & 0.252489 & 0.113 & 5.647 &  $-$  &  $-$   &  $-$  &  $-$     \\
VVV J2710857.12-410712.8 & 0.285146 & 0.124 & 4.937 & 0.066 & $-$7.98  & 0.057 & $-$13.387  \\
VVV J2704051.15-420503.6 & 0.302654 & 0.076 & 0.275 &  $-$  &  $-$   &  $-$  &  $-$     \\
VVV J2705404.77-422910.0 & 0.322360 & 0.055 & 5.595 &  $-$  &  $-$   &  $-$  &  $-$     \\
VVV J2713458.02-412608.5 & 0.330849 & 0.060 & 0.422 &  $-$  &  $-$   &  $-$  &  $-$     \\
VVV J2702945.91-413239.9 & 0.332772 & 0.019 & 5.313 & 0.236 & $-$4.388 &  $-$  &  $-$     \\
VVV J2711142.24-411611.0 & 0.358777 & 0.080 & 3.571 & 0.255 & $-$2.865 &  $-$  &  $-$     \\
VVV J2702329.86-415953.8 & 0.359989 & 0.105 & 0.209 & 0.409 &  5.568 &  $-$  &  $-$     \\
VVV J2711803.26-411635.8 & 0.370611 & 0.044 & 1.560 &  $-$  &  $-$   &  $-$  &  $-$     \\
VVV J2710200.43-413446.0 & 0.378936 & 0.019 & 5.662 & 0.220 & $-$9.100 &  $-$  &  $-$     \\ [0.5ex]
\hline\hline
\end{tabular}
}
\label{tbl:Fourier}
\end{table*}

\subsection{RR Lyr stars: reddening estimation}
\label{sec:reddening}

Generally, the reddening  can be  expressed as  the difference  of the measured and intrinsic color between two filters,

\begin{equation}
E(J-K_{\rm{s}})= (J-K_{\rm{s}}) - (J-K_{\rm{s}})_{0},
\end{equation}

in this case for $J$  and $K_{\rm{s}}$. The first term of the equation is limited by the phase  coverage and the number of points of
each  variable in  multiple  wavelengths. The  VVV,  as a  variability survey,    provides    time-series      exclusively    in
$K_{\rm{s}}$ band. For  $J$ and  $H$ band, only one  observation has been  taken during the  first year of operation, which limits the
precision  of ($J-K_{\rm{s}}$)  and ($H-K_{\rm{s}}$).  The   colors ($J-H$) and ($H-K_{\rm s}$) were calculated using the faintest
point  of the  light curve according  to  the  Fourier fit,  and   taking into  account the  ratio of amplitudes  between $J$,  $H$ and
$K_{\rm s}$ described in \cite{feast08}, as follows:

\begin{equation}
\Delta H = +0.11 + 1.65( \Delta K_{\rm s} - 0.18), \\
\end{equation}
\begin{equation}
\Delta J = -0.02 + 3.60( \Delta K_{\rm s} - 0.18). \\
\end{equation}

This approximation has a larger error for the shorter-period RRab (higher amplitudes) with  a maximum of $\pm 0.2$~mag and $\pm 0.1$~mag  
for the  $J$ and $H$ band, respectively.   However, for $Z$ and $Y$ bands (also  available in the VVV Survey), we  are unable to conduct
the  same analysis  because we lack detailed  near-IR variability studies. This problem  can be solved in the future  with a database of
well-defined  near-IR  template light-curves  that  the VVV  Templates Project \citep{catelan11b,  angeloni14} is collecting.   The next step
was   to  calculate  the   intrinsic  color   of  the   sample, that is $(J-K_{\rm{s}})_{0}$,  analogous  to  the difference  in
absolute magnitude of the  two filters. This modifies the latter reddening equation into

\begin{equation}
E(J-K_{\rm{s}}) = (J-K_{\rm{s}}) - (J-K_{\rm{s}})_0 = (J-K_{\rm{s}}) - (M_{\text{J}}-M_{\text{K}_{\rm{s}}})
\end{equation}
for  $J$  and   $K_{\rm{s}}$ bands.   The  absolute  magnitudes  were calculated using  the $P-L$ relation  given in \cite{alonso14}, 

\begin{equation}
M_{\rm{J}} = - 0.6365 - 2.347 \log{P} + 0.1747 \log{Z},
\end{equation}
\begin{equation}
M_{\rm{K}_{\rm s}} = - 0.6365 - 2.347 \log{P} + 0.1747 \log{Z},
\end{equation}
with $\sigma_{\rm J} = 0.17$~mag and $\sigma_{\rm K_{\rm s}} = 0.13$~mag the systematic errors in $J$ and $K_{\rm s}$ bands, respectively.
The errors were calculated as the difference between the average magnitudes, published by \cite{delprincipe05}, and the theoretical values obtained using 
the $P-L$ relations listed above. The theoretical propagated error of $(J-K_{\rm s})_0$ is $\sim 0.21$~mag, which introduces a difference of 0.14~mag in the distance moduli.
These relations are crucial for determining the distances to our RR~Lyr stars, because it is a very accurate solution 
to calculate the reddening corrected colors, especially in the near-IR \citep{catelan04}. 
We used the same approach as \cite{pietrukowicz12} to calculate the $\log Z = [Fe/H] - 1.765$, based on a solar   
metallicity of $Z_\odot = 0.017$ \citep{catelan04}. 
This relation was established for fundamental-mode pulsators, therefore we must exclude the RRc stars of future analyses.
Finally, these colors give us an estimate of the individual reddening of the RRab stars, which is shown in Table \ref{tbl:ead}.
We compared these reddening values with those obtained by \cite{kunder08} in the same Galactic latitude 
as our values $(\ell,b \sim -5^\circ,-9^\circ$) from the Galactic bulge fields of the MACHO survey. The mean converted value of $E(J-K_{\rm s}$)
at that latitude is $\sim 0.20$~mag, with a dispersion of $0.04$~mag, which is consistent with our measurements within the errors.
In some cases, our estimates are slightly higher than those of \cite{kunder08}, which may be due to larger distances or larger errors in the determination of 
the color $(J-K_{\rm s})$ in our light curves.

\subsection{Distance to the RR Lyrae stars}
\label{distance}

We calculated the individual  distances to the RRab stars using the $P-L$  relations  for the  $K_{\rm{s}}$ band.   This  distance can  be
expressed  in  terms  of  the absolute  and  the  extinction-corrected $K_{\rm{s}}$-band magnitudes,

\begin{equation}
\log R = 1 + 0.2(K_{\rm{ s,0 }} - M_{\rm{K}_{\rm s}} ),
\end{equation}

with  $R$ the individual  distance in pc  to our RRab  stars and converting    the   color   excess    into   extinction    using   the
law of \cite{cardelli89},

\begin{equation} 
A_{K \rm{s}} = 0.73 E(J-K_{\rm{s}}).
\end{equation}

\begin{figure}
 \centering
 \includegraphics[scale=0.46]{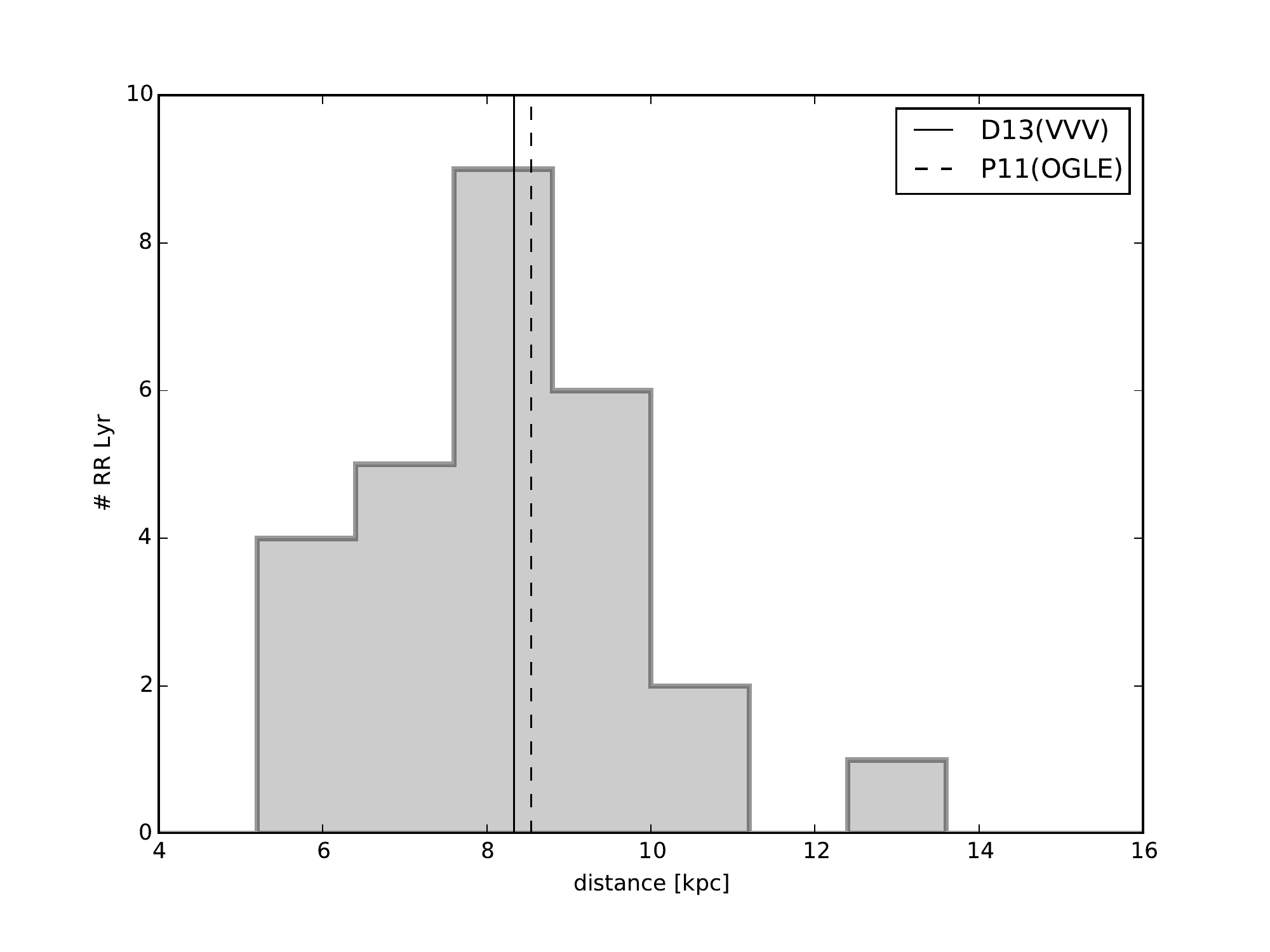}
 \label{dist}
 \caption{Distance  to  the  sample  of  RR  Lyr  stars  in   tile    \textit{b201}  in a  1.2~kpc  binning size.   The  solid (D13)  and
   dashed  lines (P11)  represent the  values of  the  Galactic center    distance   given  in  \cite{dekany13}   and  \cite{pietrukowicz12}.}
\label{fig:dist}
\end{figure}

Figure~\ref{fig:dist} shows the  histogram of the distances calculated for the individual RR~Lyr. A strong agreement with previous results is
recovered even at this high latitude ($\ell \sim -10^\circ$) with more than  the  50\%  of  the  variables in  the  central  kiloparsec  
according to  a Gaussian fit with  $d_{\rm{mean}} = 8.14$  kpc and a   dispersion   of   $\sigma_{d}    =   1.56$   kpc.    Adopting   the
reddening law of \cite{nishiyama09} does not  significantly  change the distance distribution  of our RRab stars; it adds 0.4  kpc to the
mean, but keeps the dispersion unchanged.

\begin{figure*}
 \centering
 \includegraphics[scale=0.43]{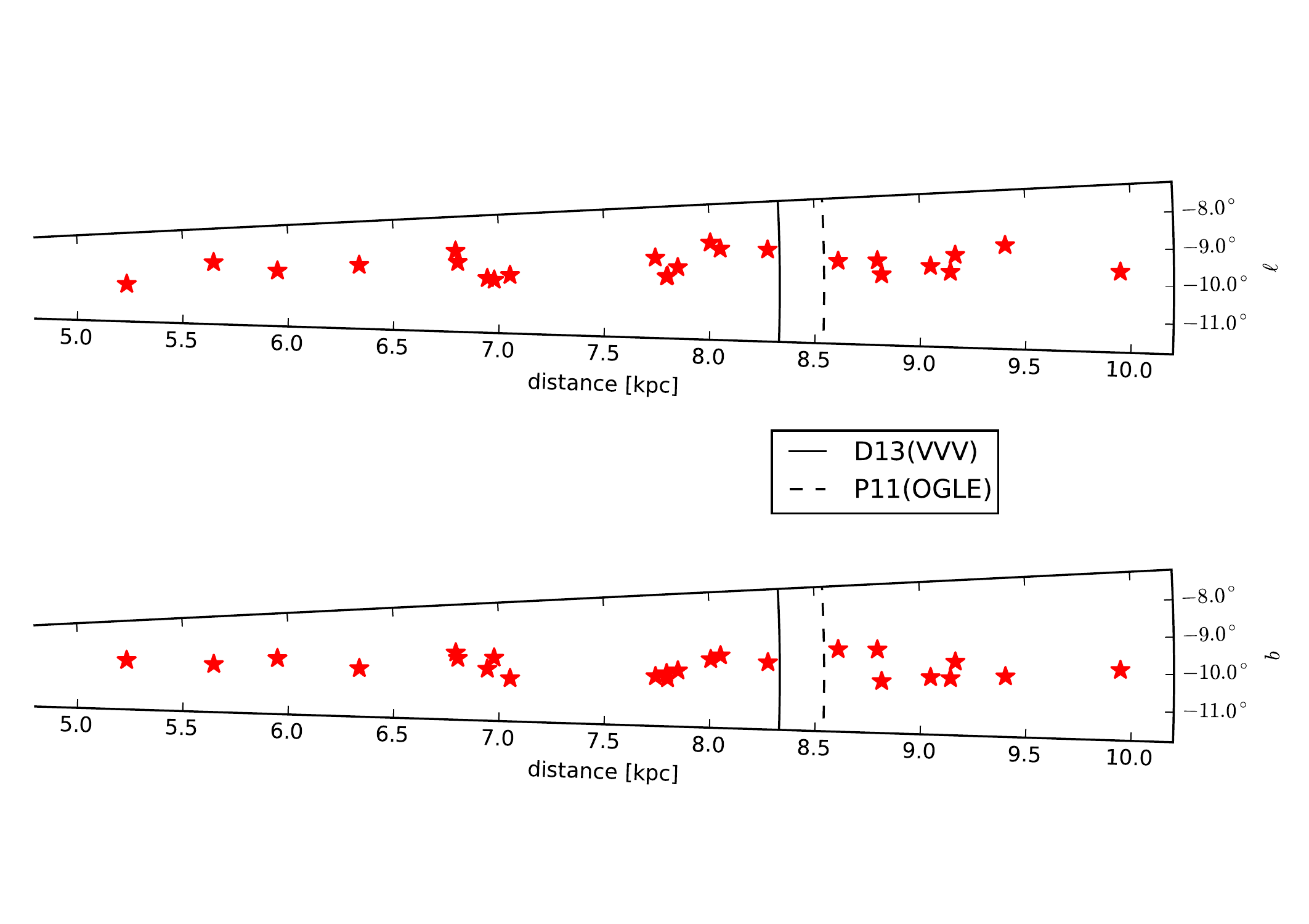} \includegraphics[scale=0.4]{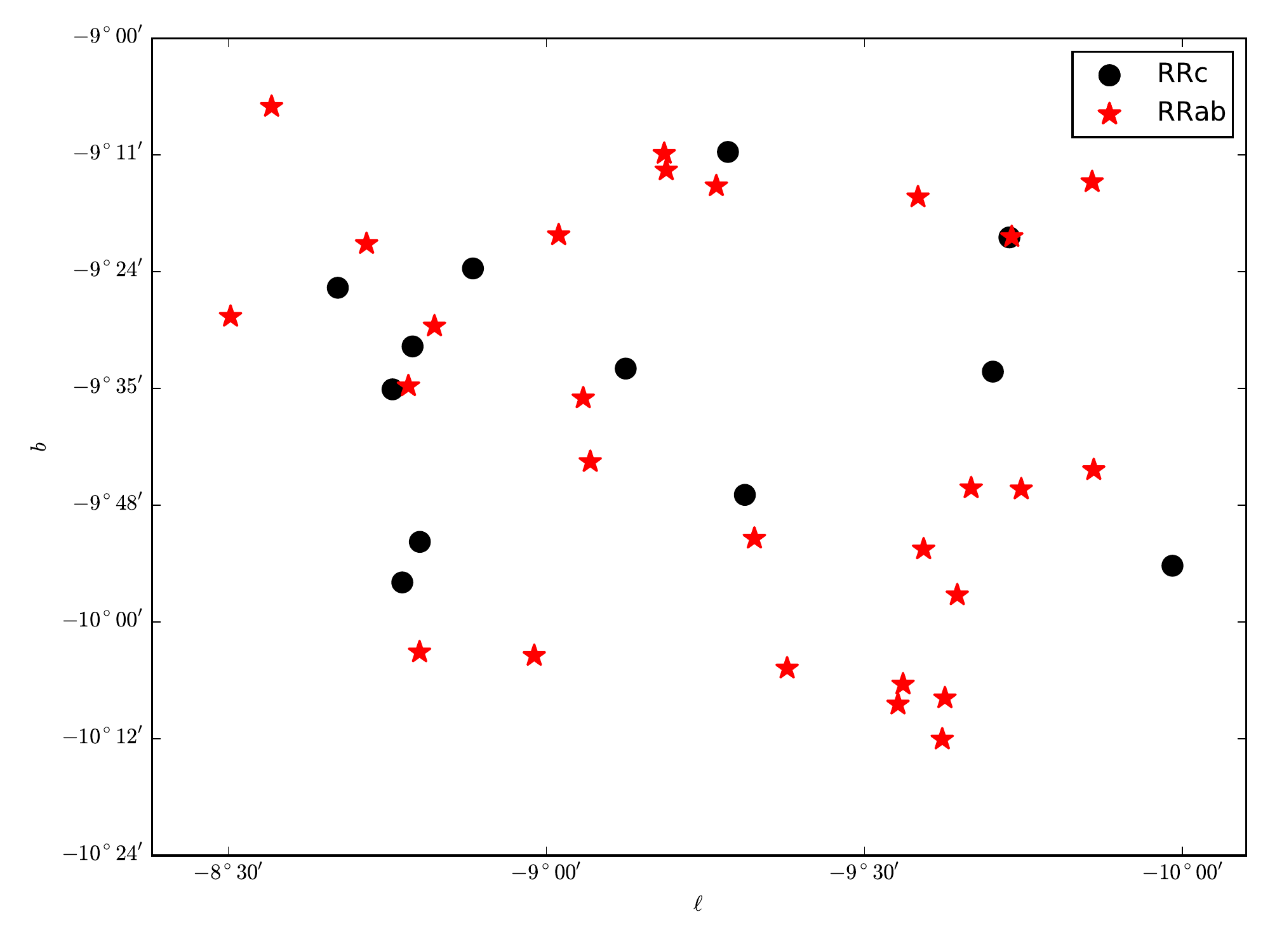}
 \label{cone}
 \caption{\textbf{Left:}  Distance  to the  sample  of  RR Lyr  stars, between  5 and  10~kpc,  in a  cone  view.  We  did  not find  any
   apparent cluster or group in the  RRab due to streams, except for the variables in the  Galactic center. The solid and  dashed lines have
   the  same meaning  as in  Fig.  \ref{fig:dist}.  \textbf{Right:} Spatial distribution in Galactic coordinates of the 39 RR~Lyr found
   in tile \textit{b201}. }
\label{fig:cone}
\end{figure*}

The  results of  this  analysis, including  reddening, extinction,  and distances with  their error are shown  in Table \ref{tbl:ead}.
The uncertainties consider the error propagation on the color, reddening,  extinction, and  distance of each  individual RRab
star.  Another  visualization  of  the  derived  distances, assuming  a constant  $\ell$ and  $b$ coordinate,  is shown  in Fig~\ref{fig:cone} 
as  a cone  view. This  approach is  useful in identifying groups or  streams of variable stars in  our FoV, which
is not  possible with their spatial  distribution ($\ell,b$) alone, as also shown in Fig.~\ref{fig:cone}. Although there may be a 
group of RR~ab stars at $d \sim 7$ kpc, these stars do not cluster in any other parameters, for example, in the period-amplitude diagram or spatially,
and hence we are hesitant to consider them as a group.

\begin{table*}
\centering
\caption{Reddening, extinction  and distances computed for  the RR~Lyr found in tile \textit{b201}. }
\begin{tabular}{c|c|c|c}
\hline \hline
VVV ID (RRab) & $E(J-K_{\rm{s}}$) & $A_{K\rm{s}}$ & Distance \\
         &     (mag)         &         (mag)         &   (kpc)  \\
\hline\hline
VVV J2712426.16-421241.6 & 0.33 $\pm$ 0.13 & 0.23 $\pm$ 0.09 & 9.14  $\pm$ 0.31 \\
VVV J2712711.57-421314.0 & 0.47 $\pm$ 0.11 & 0.32 $\pm$ 0.08 & 7.05  $\pm$ 0.21 \\
VVV J2712912.05-420223.7 & 0.18 $\pm$ 0.16 & 0.12 $\pm$ 0.11 & 9.05  $\pm$ 0.40 \\
VVV J2711512.06-415330.1 & 0.25 $\pm$ 0.13 & 0.17 $\pm$ 0.09 & 7.85  $\pm$ 0.26 \\
VVV J2703434.23-413322.1 & 0.30 $\pm$ 0.13 & 0.21 $\pm$ 0.09 & 5.95  $\pm$ 0.19 \\
VVV J2703536.01-412829.4 & 0.13 $\pm$ 0.13 & 0.09 $\pm$ 0.09 & 8.79  $\pm$ 0.28 \\
VVV J2703343.00-412731.0 & 0.38 $\pm$ 0.13 & 0.26 $\pm$ 0.09 & 8.61  $\pm$ 0.31 \\
VVV J2711523.24-413619.8 & 0.42 $\pm$ 0.13 & 0.29 $\pm$ 0.09 & 6.34  $\pm$ 0.22 \\
VVV J2710804.03-413241.7 & 0.28 $\pm$ 0.14 & 0.19 $\pm$ 0.09 & 9.16  $\pm$ 0.33 \\
VVV J2704944.51-412248.6 & 0.04 $\pm$ 0.16 & 0.03 $\pm$ 0.11 & 6.80  $\pm$ 0.30 \\
VVV J2714215.15-414058.6 & 0.43 $\pm$ 0.15 & 0.30 $\pm$ 0.10 & 7.74  $\pm$ 0.32 \\
VVV J2710759.32-411700.5 & 0.19 $\pm$ 0.14 & 0.13 $\pm$ 0.09 & 5.64  $\pm$ 0.21 \\
VVV J2714823.59-413121.7 & 0.05 $\pm$ 0.14 & 0.04 $\pm$ 0.09 & 9.40  $\pm$ 0.35 \\
VVV J2711644.06-411744.6 & 0.19 $\pm$ 0.14 & 0.13 $\pm$ 0.10 & 8.27  $\pm$ 0.31 \\
VVV J2710158.99-410727.3 & 0.21 $\pm$ 0.10 & 0.14 $\pm$ 0.07 & 8.05  $\pm$ 0.20 \\
VVV J2704656.26-421805.3 & 0.37 $\pm$ 0.09 & 0.25 $\pm$ 0.06 & 13.58 $\pm$ 0.28 \\
VVV J2701149.18-420358.3 & 0.34 $\pm$ 0.11 & 0.23 $\pm$ 0.08 & 5.23  $\pm$ 0.15 \\
VVV J2705101.04-405300.9 & 0.17 $\pm$ 0.12 & 0.12 $\pm$ 0.08 & 6.79  $\pm$ 0.21 \\
VVV J2711836.69-405946.8 & 0.34 $\pm$ 0.13 & 0.23 $\pm$ 0.09 & 8.00  $\pm$ 0.28 \\
VVV J2705334.00-421303.7 & 0.45 $\pm$ 0.10 & 0.31 $\pm$ 0.07 & 10.93 $\pm$ 0.26 \\
VVV J2702317.06-420003.9 & 0.33 $\pm$ 0.10 & 0.22 $\pm$ 0.07 & 6.98  $\pm$ 0.17 \\
VVV J2705622.48-420853.7 & 0.45 $\pm$ 0.12 & 0.31 $\pm$ 0.09 & 6.94  $\pm$ 0.23 \\
VVV J2711018.90-421253.4 & 0.34 $\pm$ 0.12 & 0.24 $\pm$ 0.08 & 7.79  $\pm$ 0.23 \\
VVV J2712342.04-421647.8 & 0.08 $\pm$ 0.14 & 0.06 $\pm$ 0.09 & 7.80  $\pm$ 0.29 \\
VVV J2712855.63-421832.3 & 0.20 $\pm$ 0.09 & 0.13 $\pm$ 0.07 & 8.81  $\pm$ 0.21 \\
VVV J2710638.97-420755.7 & 0.20 $\pm$ 0.13 & 0.14 $\pm$ 0.09 & 9.95  $\pm$ 0.35 \\
VVV J2702400.61-415026.5 & 0.11 $\pm$ 0.15 & 0.08 $\pm$ 0.10 & 10.83 $\pm$ 0.45 \\ [0.5ex]
\hline\hline
\end{tabular}
\label{tbl:ead}
\end{table*}

An  interesting topic  of the  RR~Lyr studied  in large  areas  is the possibility  that   they  can  trace  an   old  stream-like  structure
\citep{sesar13}   in addition to the   spheroidal   shape  described   by \cite{dekany13}. Unfortunately, the small number  of variables in
this single tile prevent us  from making any strong statements about   the presence of streams in this direction.

\section{Summary}
\label{sec:conclusion}

A search for RR~Lyr stars was performed in the VVV tile \textit{b201}, one of the outermost bulge regions  covered by the VVV Survey.  A total
of 39 RR Lyr were found, of which  27 and 12 belong to the ab- and c-type, respectively.  For the  total sample, coordinates, periods, amplitudes,
and  near-IR  colors   were  presented.   Using  the   International Variable  Star Index, we found counterparts  of the ten
previously  known  RR  Lyr  in  the  field,  in  agreement  with  past measurements. We found differences in only  three of them, for which we
suggested  a revised  period.  From  the analysis  of  the RRab sample,  individual  reddening,  extinction, and  distance  were
obtained,  based   on  line-of-sight  extinction   values  given  by \cite{cardelli89}  and  the   $P-L$  relation  of  \cite{alonso14}.
The distribution  of these variables on  the Milky Way follows a  central distribution around $\sim 8.1$ and 
$\sim 8.5$ kpc. using the Cardelli or Nishiyama extinction law, respectively. This value is consistent with the
results of \cite{dekany13}, placing the center of the distribution at $\sim$8.3 kpc. An analysis  of a larger area around 
$\ell \sim -10^\circ$ will be conducted to complement the studies of the inner Galactic bulge that are already published.

\begin{acknowledgements}
We thank the anonymous referee for comments that helped improve the presentation of our results.
We gratefully acknowledge  the use of data from  the ESO Public Survey program ID  179.B-2002 taken with  the VISTA telescope,  data products
from the Cambridge Astronomical  Survey Unit.  Support for the authors is provided by  the BASAL CATA Center for  Astrophysics and Associated
Technologies through  grant PFB-06, and the Ministry  for the Economy, Development,  and Tourism's  Programa Iniciativa  Cient\'ifica Milenio
through grant IC120009, awarded to Millenium Institute of Astrophysics (MAS), FONDECYT  Regular grants No. 1130196 (D.M.)  and 1141141 (M.C.,
C.N.,  F.G.).  F.G.  and  C.N. acknowledge  support from  CONICYT-PCHA Mag\'ister  Nacional  2014-22141509  and 2012-22121934,  respectively.
R.C.R. acknowledges support by Proyecto Fondecyt Postdoctoral \#3130320.
R.K.S.   acknowledges  support   from  CNPq/Brazil   through  projects 310636/2013-2 and 481468/2013-7.  We gratefully acknowledge the use of
IPython \citep{perez07}, TOPCAT  \citep{taylor05}, and the ALADIN sky atlas \citep{bonnarel00}  in this  research and  the  International Variable
Star   Index   (VSX)   database,   operated   at   AAVSO,   Cambridge, Massachusetts, USA.
\end{acknowledgements}

\bibliography{biblio}

\end{document}